\documentclass[12pt]{elsart}
\usepackage{epsf}
\usepackage{amssymb}

\newcommand{\PSfigure}[2]{%
    \centerline{\epsfxsize=#2\textwidth\epsffile{#1}}
}

\begin{document}
\bibliographystyle{elsart-num}

\begin{frontmatter}

\title{Propagation of waves in metallic photonic crystals at low frequencies and some theoretical aspects of left-handed materials}

\author{A. L. Pokrovsky},
\author{A. L. Efros},

\address{
    Department of Physics \\
    University of Utah\\
    Salt Lake City, UT 84112 \\
    USA
}

\thanks{%
  We are grateful to A. Zakhidov and V. Vardeny who have attracted our
  attention to this interesting problem and to L. P. Pitaevskii and  B. L. Spivak
  for important discussions. 
  This work supported by the NSF grant DMR-0102964.
}


\begin{abstract}
An analytical theory of low frequency electromagnetic waves in 
metallic photonic crystals with a small volume fraction of a metal 
is presented.
The evidence of the existence of such waves has been found recently 
via experiments and computations.
We have obtained an exact dispersion equation for $\omega (k)$ and studied 
the cutoff frequency $\omega_0 = \omega (0)$ as a function of 
parameters of the photonic crystal.
An analytical expression for 
the permittivity  $\epsilon$ is calculated.
It is shown, that if the crystal is embedded
into a medium with negative $\mu$, it has no
propagating modes at any frequency. 
Thus, such a compound system is not a left-handed material (LHM). 
The recent experimental results on the LHM are discussed.
\end{abstract}

\end{frontmatter}

Veselago has shown\cite{ve} that if in some frequency range
both permittivity $\epsilon$ and permeability $\mu$ are negative,
the electromagnetic waves (EMW's) propagate
but they have some peculiar properties, from which the primary one is that
vectors {\bf k}, {\bf E}, {\bf H} form rather a left-handed 
than a right-handed triple of vectors. 
Materials, with this property are named left-handed materials (LHM).

The San Diego group\cite{sm3} uses metallic photonic crystals (MPC's) as
a technological base for the LHM in the GHz frequency range. 
This idea comes from the pioneering computational
 and experimental studies of a few
 groups\cite{yab,soc1,soc,p3,sm2} which have found very low frequency EMW's 
in the MPC.
These EMW's propagate
above a very low cutoff frequency for which various groups obtained different values.
Since the parameters of the MPC's were also different, it was not clear 
whether or not the physics of these modes is the same.
The suggested interpretations are very controversial.
The group of Soukoulis\cite{soc} qualitatively interpreted the effect 
of propagation in terms of waveguide modes, 
while the group of Pendry\cite{p3} presented a completely original 
physical picture based upon a new longitudinal mode, called ``plasma mode".
According to Pendry {\it et al.}\cite{p3} 
the resulting effective permittivity has a plasma-like behavior
\begin{equation}
\label{eps}
\frac{\epsilon}{\epsilon_0}= 1-\frac{\omega_p^2}{\omega(\omega+i\Gamma)},
\end{equation}
however, expression for the ``plasma" frequency contains the velocity of light and has a form
\begin{equation}
\label{om}
\omega_p^2=\frac{2 \pi c^2}{d^2 \ln (d/R)},
\Gamma=\frac{\epsilon_0d^2\omega_p^2}{\pi R^2\sigma},
\end{equation}
where $d$ is the lattice constant, $R$ is the radius of the metallic wires,
$\sigma$ is the static conductivity of the metal. The same results for
$\epsilon$ and $\omega_p$ have been later obtained 
theoretically by Sarychev and Shalaev\cite{sar}.

The San Diego group accepted the ``plasma model" and considered the
negative $\epsilon$ at $\omega<\omega_p$ as one of the two crucial 
conditions for creation of the LHM. 
They have reported the
first observation of the anomalous transmission and negative refraction 
in a compound system of split ring
resonators (SRR's) and MPC\cite{sm3,sm}. 
According to Pendry {\it et al.}\cite{p1} such 
resonators create a negative bulk magnetic
permeability due to the anomalous dispersion.


In this paper we present an analytical theory of the 
EMW's propagating in the MPC without SRR's at very low frequencies.
The theory is based upon the parameters $f \ll 1$ and $\omega R/c \ll 1$, 
where $f$ is the volume
fraction of a metal in the MPC.
First we derive an exact dispersion equation for the s-polarized
EMW in the system of infinite parallel thin straight wires
ordered in a square lattice. 
The electric field of the wave is directed along
the wires (z-axis), while the wave vector ${\bf k}$ is in the $x$-$y$
plane.  Assume that the total current in each wire is $I_0\exp[
-i(\omega t -{\bf k\cdot r}_i)]$, where ${\bf r}_i$ is a 
 two-dimensional radius-vector of the wire in the $x$-$y$ plane.  The
 external solution for electric field $E_z$ of one wire, located at ${\bf
 r}_i=0$, has a form
\begin{equation}
E_{z}^i= \frac{I_0 \mu_0 \omega}{4} H^{(2)}_0\left(\frac{\omega}{c} \rho \right),
\end{equation}
where $H^{(2)}_0=J_0-iN_0$ is the Hankel function which decays
exponentially at ${\rm Im}\omega <0$. Neither $J_0$ nor $N_0$ has
this important property.  Here and below we omit the time dependent factor.
The solution is written in cylindrical coordinates $z,\rho,\phi$ and
it satisfies the boundary condition $B_{\phi}=(i/ \omega)dE_z/ d\rho=
I_0\mu_0/ 2\pi \rho$ at $\rho = R$.

The electric field created by all wires is
\begin{equation}
\label{e}
E_z({\bf r})=\frac{I_0 \mu_0 \omega}{4}e^{i{\bf k\cdot r}}
        \sum_{j}e^{i{{\bf k\cdot} ({\bf r}_j-{\bf r})}}
        H^{(2)}_0(\frac{\omega}{c} \rho_j),
\end{equation}
where $\rho_j = \sqrt{(x-x_j)^2+(y-y_j)^2}$,
summation is over all sites of the square lattice and the sum is
a periodic function of ${\bf r}$.
The dispersion equation follows from the boundary condition\cite{lan} 
that relates the total electric field at the surface of any wire $l$ 
to the total current through the wire $E_z(R)=I_0 \exp{(i {\bf k}\cdot{\bf r}_l)}/\sigma_{ef} \pi R^2$, 
where $\sigma_{ef}=2\sigma J_1(\kappa R) / \kappa R J_0(\kappa R)$,
$\kappa=(1+i)/\delta$, and $\delta$ is
 the skin depth. At small frequencies, when $\delta > R$ one gets
 $\sigma_{ef}\approx \sigma$. At high frequencies, when $\delta \le
 0.1 R$, one gets the Rayleigh formula $\sigma_{ef}\approx (1+i)\sigma
 \delta/R$.

Note, that the EMW exists mostly if the skin-effect in the wires is
strong.  Using Eqs.(\ref{eps},\ref{om}) one can show that $\Gamma
/\omega_p = \delta^2 / (R^2 \ln{d/R})$, where $\delta$ is the skin
depth at $\omega = \omega_p$ (see also Ref.\cite{shal}).  
Thus, with the logarithmic accuracy one can say that $\Gamma \ll \omega_p$ if
$\delta \ll R$.
We show below that the exact solution has similar properties.

Finally, the dispersion equation for
$\omega({\bf k}) \equiv (\chi - i \gamma)c/d$ has a form
\begin{equation}
\label{dsp}
(\chi - i \gamma) \sum_{l,m} e^{i d( k_x l+ k_y m)} H_0^{(2)} ( z_{lm}) =
 \frac{4 c \epsilon_0}{d \sigma_{ef} f},
\end{equation}
where where $z_{lm} =  (\chi - i \gamma) \sqrt{l^2 + m^2 + (R/d)^2}$, $l$ and $m$ 
are integer numbers and
the small term $(R/d)^2$ under the square root is important
 only when $l = m = 0$.  Taking real and imaginary parts of Eq.(\ref{dsp})
one gets two equations for $\chi$ and $\gamma$.



\begin{figure}
    \PSfigure{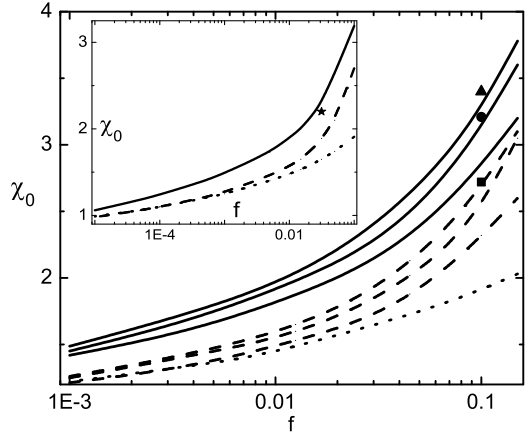}{0.5}
    \caption{%
    The real part of the dimensionless cutoff frequency
       	$\chi_0$ as a function of the
        volume fraction of metal $f$.
	Solid, dashed, and dotted lines represent solutions 
	of Eq.(\ref{dsp0}), Eq.(\ref{dsp_a}), Eq.(\ref{om}) respectively.
	The experimental data of Ref.\protect\cite{sm}($\bigstar$)
	are shown together with numerical data of
	Ref.\protect\cite{soc} ($\blacktriangle$,
	$\bullet$, $\blacksquare$). 
	On the main plot
 	$\alpha = 0.024$, $d = 12.7 \mu m$ for upper solid, upper dashed lines,
 	and the point  $\blacktriangle$;
 	$\alpha = 0.078$, $d = 1.27 \mu m$ for middle solid, middle
        dashed lines, and the point $\bullet$;
        $\alpha = 0.246$, $d = 0.13 \mu m$ for lower solid,
	lower dashed lines, and the point $\blacksquare$.
        On the insert $\alpha = 3.4\cdot 10^{-4}$, $d = 8.0 mm$.
    }
    \label{fig}
\end{figure}

Now we study the cutoff frequency for the EMW's $\omega_0$, which is 
the solution of Eq.(\ref{dsp}) at $|{\bf k}| = 0$:
\begin{equation}
\label{dsp0}
(\chi - i \gamma) \sum_{l,m} H_0^{(2)} (z_{lm}) =
 \frac{4 c \epsilon_0}{d \sigma_{ef} f}.
\end{equation}
The numerical results for the real part of the frequency
are shown in Fig. \ref{fig}.
One can see that $\chi$ is of the order of few units.
The values of $\gamma$ are of the order of the right hand side of 
Eq.(\ref{dsp0}).
Thus, $\chi \gg \gamma$ if $f \sigma_{ef} / \epsilon_0 \omega \gg 1$.

In addition to the numerical solution we propose an approximation valid
at a very small $f$, when $|\ln f|\gg 1$.
We separate the term with $l = m = 0$ in Eq.(\ref{dsp0}) and
 substitute the rest of the sum by the integral.  Then
\begin{equation}
\label{dsp_a}
(\chi - i \gamma) \left[ \frac{2 i}{\pi} \left( \ln{\frac{2}{\chi
        \sqrt{f/\pi}}} - {\bf C} \right) - \frac{4 i}{(\chi - i
        \gamma)^2} \right] 
        = \frac{4 c \epsilon_0}{d \sigma_{ef}(\chi) f},
\end{equation}
where we assume that $\gamma \ll \chi$.  
Here ${\bf C}$ is the Euler's constant. 

Figure \ref{fig} compares our result for the real part of the
cutoff frequency given by  Eq.(\ref{dsp0}) with the results given by
Eqs.(\ref{om},\ref{dsp_a}).  In these calculations we assume
that $\sigma_{ef}$ is given by the Rayleigh formula so that the right hand
side of Eqs.(\ref{dsp0},\ref{dsp_a}) has a form $\alpha (1-i)
\sqrt{\chi}$, where $\alpha = 2 c \epsilon_0 R/f d \sigma \delta$ and
$\delta$ is taken at $\omega = c/d$. One can see from Fig.\ref{fig} that
approximation Eq.(\ref{dsp_a}) is much better than approximation
Eq.(\ref{om}).  Both approximations coincide at small $f$ and they are
accurate at extremely low values of $f$ ($\sim 10^{-7}$) when the
logarithmic term in Eqs.(\ref{om},\ref{dsp_a}) is very large.  The
computational and experimental data of Refs.\cite{soc,sm} are also
shown at Fig.~\ref{fig} and they are in a good agreement with
Eq.(\ref{dsp0}).
The result of Pendry group \cite{p3} for the cutoff frequency is 
$8.2{\rm\ GHz}$, while Eq.(\ref{dsp0}) gives  $8.6{\rm\ GHz}$. 
Experimental result of Bayindir {\it et al.}\cite{bay} is $11.67 {\rm\ GHz}$ while our 
Eq.(\ref{dsp0}) gives $11.8{\rm\ GHz}$.
Thus, we can make a conclusion
that the San Diego group, group of Pendry and Soukoulis group
discuss the same mode but at different values of parameters  
and that our analytical theory describes the same mode as well.

\begin{figure}
    \PSfigure{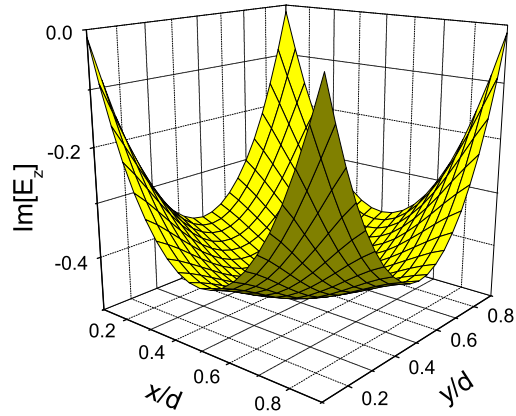}{0.5}
    \caption{%
       Imaginary part of the electric field inside the lattice cell 
       in the units of $I_0 \mu_0 c/d$. 
       The radius of the wires $R = 0.8$ mm, 
       the lattice constant $d = 8$ mm, 
       the static conductivity of the metal of the wires 
       $\sigma = 5.88\cdot 10^7$ ($\Omega\ \cdot$ m)$^{-1}$.
    }
    \label{fig01}
\end{figure}

\begin{figure}
\PSfigure{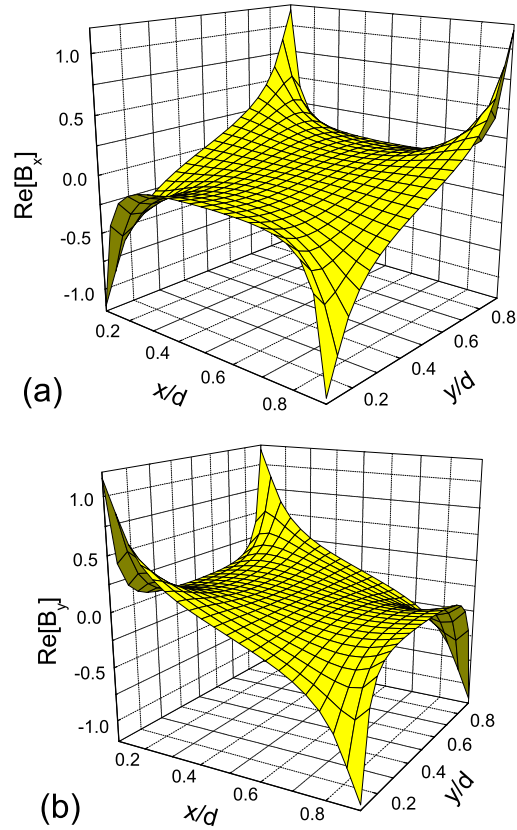}{0.5}
\caption{Real part of the $x$-($a$), and $y$-component(b) of the magnetic 
	induction ${\bf B}$ inside the lattice cell  in the units of $I_0 \mu_0/d$. 
	The radius of the wires $R = 0.8$ mm, 
	the lattice constant $d = 8$ mm, the static conductivity of the metal of the wires 
	$\sigma = 5.88\cdot 10^7$ ($\Omega\ \cdot$ m)$^{-1}$. }
\label{fig02}
\end{figure}

At Fig. \ref{fig01} we plot the imaginary part of the $z$-component
of the electric field in the unit cell as calculated from Eq.(\ref{e}) with $k = 0$.
Real parts of the $x$- and $y$- components of the magnetic induction, 
are shown in Fig. \ref{fig02}.
Real part of the $E_z$ and imaginary parts of $B_x$, $B_y$ are 
much smaller than imaginary part of $E_z$ and real part of $B_x$, $B_y$
correspondingly if $\gamma \ll 1$. 
As one can see, the fields are strongly modulated inside the 
lattice cell and $E_z$ is close to zero near each wire so that absorption is small.
This fact allows the low frequency mode propagate in the MPC under the condition 
$f \sigma_{ef} /\epsilon_0 \omega \gg 1$.

Now we find the component of electric permittivity
$\epsilon(\omega) = \epsilon_{zz}$,
which describes the s-polarized extraordinary waves
in the uniaxial crystal.
It is defined by the relation $\epsilon = \epsilon_0 + i \tilde{\sigma} d/\chi c$,
where effective macroscopic conductivity $\tilde{\sigma}$ relates
average current density $\overline{\jmath}_z$ to the average electric
field $\overline{E}_z$ by equation $\overline{\jmath}_z =
\tilde{\sigma} \overline{E}_z$.  To find $\tilde{\sigma}$ we introduce
an external electric field $\mathcal{E}_z e^{ -i \omega t}$.  
Using the boundary condition on a wire, which now has a form
\begin{equation}
\label{bc}
\frac{I_0 c (\chi - i \gamma) \mu_o}{4 d} \sum_{l,m}{}
      H_0^{(2)}(z_{lm}) + \mathcal{E}_z 
      = \frac{I_0}{\pi R^2
      \sigma_{ef}}.
\end{equation}
one can find a
relation between the current and the average field which gives both
 $\tilde{\sigma}$ and $\epsilon$. Finally one gets
\begin{equation}
\label{epsilon}
\frac{\epsilon}{\epsilon_0} = 1-\left\{ \displaystyle \frac{\chi}
{\chi - i \gamma} - i \chi \left[ \frac{\chi - i \gamma}{4} \right. \right.
\left. \left. \sum_{l,m}
H_0^{(2)}\left(z_{lm}\right) -\frac{c \epsilon_0}{f d \sigma_{eff}}
\right]\right\}^{-1}.
\end{equation}

The expression in the square brackets of Eq.(\ref{epsilon})
is the dispersion equation (\ref{dsp0}).
One can see that ${\rm Re}\epsilon$
changes sign at $\omega = \omega_0$ and becomes negative 
at $\omega<\omega_0$, where 
$\omega_0$ is the root of the dispersion equation (\ref{dsp0}).

To find $\omega({\bf k})$ one should solve  Eq.(\ref{dsp}).
For small $|{\bf k}|$ one can get an analytical result 
$\omega^2 = c^2 k^2 + \omega_0^2$,
which is isotropic in the $x$-$y$ plane.


Now we discuss the possibility 
of creation of the LHM using the negative $\epsilon$ of the MPC. 
Suppose that the wires are embedded into a medium with the
negative magnetic permeability $\mu$.  One can see that in this case
the propagation of any EMW is suppressed.  Indeed, instead of
Eq.(\ref{dsp}) one gets equation
\begin{equation}
\label{dsp_bad}
\sum_{j} e^{i ( k_x x_j+ k_y y_j)} K_0 \left( \frac{\omega}{c_0}
       \sqrt{x_j^2 + y_j^2 + R^2} \right) 
       = \frac{2 i}{|\mu| \omega R^2 \sigma_{ef} },
\end{equation}
where $c_0=1/\sqrt{|\mu| \epsilon_0}$ and $K_0$ is the modified Bessel
function.  One can see that at $|{\bf k}| = 0$ all the terms on the
left hand side of this equation are positive and real if ${\rm Im}\omega$ is small. 
Thus, if the right hand side is small, the equation cannot be satisfied.
At small values of $\omega$ and
$|{\bf k}|$ summation in the Eq.(\ref{dsp_bad}) can be substituted by
integration. Assuming that ${\rm Re}\omega \gg {\rm Im}\omega$ one
gets $1 + k^2 c_0^2/\omega^2=-i\sigma_{ef} f/\omega \epsilon_0$.
This equation does not have real solutions for $\omega (k)$.  Thus, at
negative $\mu$ there are no propagating modes at any frequency under
the study.
%

Now we compare this result with the theoretical idea\cite{sm,p2} to obtain the LHM,
where negative $\epsilon$ is created by the system of wires and 
negative $\mu$ is created in some other way.
This idea is based upon the assumption that the 
negative $\epsilon$ at $\omega< \omega_0$ results from a ``longitudinal plasma mode".
It is taken for granted that its frequency 
is independent of magnetic properties of the system, which is 
usually the case for plasmons.
However, the mode discussed above is not a plasma mode (see also \cite{pcom}). 
One can show that
this mode has zero
average value of the magnetic induction 
$\overline{{\bf B}}$ over the unit cell.
In this sense this is indeed a longitudinal mode.
But the average
value of the magnetic energy, which is 
proportional to $\overline{{\bf B}^2}$, is not zero and it is large.   
The physics of this mode is
substantially related to the magnetic energy.  That is why
negative $\mu$ completely destroys  this mode. It destroys also 
the region of negative $\epsilon$. In fact 
 this could  be
predicted
from the observation that $\omega_0 \sim c/d$ becomes
imaginary at negative $\mu$.
For example, one can see from Eqs.(\ref{eps},\ref{om}) that at $\mu<0$ one gets
$\omega_p^2<0$ and $\epsilon>0$ at all frequencies assuming that $\Gamma$ is small.

Thus, we have shown that the simple explanation \cite{p2,sm} 
of the negative refraction
in  the compound system of the MPC and SRR's, 
based upon the  permittivity $\epsilon$ of the MPC and the negative permeability $\mu$ 
of the SRR's does not work because negative $\mu$ blocks propagation of EMW's  in the MPC. 
The propagation observed by the San Diego group might be a manifestation of the remarkable conclusion
 of Landau and Lifshitz (See Ref.\cite{lan} p.268) that $\mu (\omega)$ does not have physical
meaning starting with some low frequency. Then, 
the explanation of the negative refraction in this particular 
system would be outside the simple Veselago scenario (see Ref.\cite{notomi} as an example). 
In any case, to explain the negative refraction one should use a 
microscopic equation similar to Eq.(\ref{dsp}) 
but with the SRR's included.

\bibliography{loww}

\end{document}